# Crystal orientation dictated epitaxy of ultrawide bandgap 5.4 - 8.6 eV α-(AlGa)$_2$O$_3$ on m-plane sapphire

**Authors**

Riena Jinno[1*†,] Celesta S. Chang,[2,3*] Takeyoshi Onuma,[4] Yongjin Cho,[1] Shao-Ting Ho,[5] Michael C. Cao,[3] Kevin Lee,[1] Vladimir Protasenko,[1] Darrell G. Schlom,[5] David A. Muller,[3,6] Huili G. Xing,[1,5] and Debdeep Jena[1,5‡,]

**Affiliations**

[1] *School of Electrical and Computer Engineering, Cornell University, Ithaca, New York 14853, USA*

[2] *Department of Physics, Cornell University, Ithaca, New York 14853, USA.*

[3] *School of Applied and Engineering Physics, Cornell University, Ithaca, New York 14853, USA.*

[4] *Department of Applied Physics, Kogakuin University, 2665-1 Hachioji, Tokyo 192-0015, Japan.*

[5] *Department of Material Science and Engineering, Cornell University, Ithaca, New York 14853, USA.*

[6] *Kavli Institute for Nanoscale Science, Cornell University, Ithaca, New York 14853, USA.*

**Abstract**

Ultra-wide bandgap semiconductors are ushering in the next generation of high power electronics. The correct crystal orientation can make or break successful epitaxy of such semiconductors. Here it is discovered that single-crystalline layers of α-(AlGa)$_2$O$_3$ alloys spanning bandgaps of 5.4–8.6 eV can be grown by molecular beam epitaxy. The key step is found to be the use of *m*-plane sapphire crystal. The phase transition of the epitaxial layers from the α- to the narrower bandgap β-phase is catalyzed by the *c*-plane of the crystal. Because the *c*-plane is orthogonal to the growth front of the *m*-plane surface of the crystal, the narrower bandgap pathways are eliminated, revealing a route to much wider bandgap materials with structural purity. The resulting energy bandgaps of the epitaxial layers span a range beyond the reach of all other semiconductor families, heralding the successful epitaxial stabilization of the largest bandgap materials family to date.

---

[*] These authors contributed equally to this work.
[†] Present address: Faculty of Pure and Applied Physics, University of Tsukuba, Tsukuba 305-8571 Japan.
[‡] Corresponding author. Email: djena@cornell.edu

**Introduction**

Since the middle of the last century, semiconductor materials have made a steady climb to the top of the periodic table to wider energy bandgaps. Starting from the early days of Ge, to the Si revolution, and the current wide-bandgap SiC and GaN materials, each new generation of semiconductors have enabled electronic, photonic, and sensing and actuation functions in regimes that were considered impossible before. Next generation electronics and photonics needs much larger bandgaps beyond the established material families. Ultra-wide bandgap (UWBG) semiconductor materials for power and microwave electronics and deep-UV photonics include AlN, BN, diamond, and $Ga_2O_3$ (*1–4*). The recent availability of single-crystal β-$Ga_2O_3$ substrates has generated significant interest in this wide bandgap semiconductor material family for high-voltage electronics and UV photonics (*5, 6*). Efforts to create β-$(AlGa)_2O_3$ with high Al contents to obtain ultra-wide bandgaps to exceed the bandgap of ~6 eV available in the nitride material family in AlN have met a roadblock. Because β-$(AlGa)_2O_3$ is not the energetically favored crystalline phase for large Al compositions, the crystal converts to competing structural phases when grown on β-$Ga_2O_3$ substrates (*7–10*).

This leads to the intriguing question: can single phase and highly crystalline α-$Al_2O_3$ of bandgap $E_g$~8.8 eV, (*11*) and its alloys and heterostructures with α-$Ga_2O_3$ of bandgap $E_g$~5.3 eV (*12–14*) be grown directly on α-$Al_2O_3$ (sapphire) substrates? If this can be done, *and* if these layers can be controllably doped, it will open new application arenas. The ultra-wide bandgap energies (~5.3–8.8 eV), a large portion of which is unachievable in the nitride semiconductor family, and the attractive dielectric constants (~11 for || *c*-axis and ~9 for ⊥ *c*-axis for α-$Al_2O_3$) (*15, 16*) offer the possibility to take semiconductor electronics and photonics into regimes that currently remain out of reach. In this work, we report that the first of these challenges can indeed be met: high crystalline quality, single phase α-$(AlGa)_2O_3$ can be grown epitaxially directly on *m*-plane sapphire, spanning bandgaps ~5.3–8.8 eV. i.e. the entire Al composition of the alloy from 0 to 1. Though the conductivity control of all compositions is not achieved over the entire range of bandgaps yet, initial indications point that this is feasible. The bandgaps achieved put this material family significantly beyond those that are available today, and the correct cost-effective, large-area substrates for achieving this are already available.

**SIGNIFICANCE OF CRYSTAL ORIENTATION**

The orientation of the substrate crystal determines the subsequent crystal phases in unique ways in this UWBG material family. Homoepitaxial growth of α-$Al_2O_3$ films has been reported on *c*-, *a*-, and *r*-plane substrates using molecular beam epitaxy (MBE) (*17, 18*), and pulsed-laser deposition (PLD) (*19, 20*), but no study of their optical properties exist. On the other end of the α-$(Al_xGa_{1-x})_2O_3$ alloy system, single-crystalline α-$Ga_2O_3$ films and conductive, n-type doped films are successfully grown on sapphire substrates by mist chemical vapor deposition (CVD) (*12, 21*) and halide vapor phase epitaxy (HVPE) (*22, 23*). α-$(AlGa)_2O_3$ alloys have also been grown using MBE, [27] mist-CVD (*25–28*), and PLD (*20*). Among the growth methods, MBE offers the capability to maintain sharp interfaces across layers in heterostructures of different alloy compositions (*8*, *18*, *29*, *30*). MBE growth of α-(Al$_x$Ga$_{1-x}$)$_2$O$_3$ of high Ga content is hampered because metastable α-Ga$_2$O$_3$ in the rhombohedral corundum structure has a propensity to revert to the thermodynamically most stable monoclinic β-Ga$_2$O$_3$, especially whenever a corundum *c*-plane becomes available. This can occur either on the sapphire substrate itself, or in the underlying corundum structure of the epitaxial layers. Preventing this phase transformation holds the key to access to the desired UWBG family, as indicated in Fig. 1.

The growth of β-Ga$_2$O$_3$ on *c*-plane sapphire substrates has been found to have the epitaxial relationship of ($\bar{2}$01) β-Ga$_2$O$_3$ ∥ (0001) sapphire (*31–33*). Recent studies by scanning transmission electron microscopy (STEM) have revealed that when grown on *c*-plane sapphire, a three-monolayer-thick coherent α-Ga$_2$O$_3$ first forms, and the subsequent epitaxial layers undergo a crystalline phase transition to the β-phase (*32*), as shown in Fig. 1A. Also shown in the figure are the planes ($\bar{2}$01) β-Ga$_2$O$_3$ ∥ (0001) sapphire, which cause this transition. As shown in Fig. 1A, if the hexagonal sapphire crystal plane is rotated from the *c*-plane to the *r*-plane, the *c*-plane is now at an angle, and can potentially stabilize the growth of α-Ga$_2$O$_3$ by avoiding the crystalline phase transition. In 2018, Kracht *et al.* attempted this strategy: they reported the growth of α-Ga$_2$O$_3$ on *r*-plane sapphire substrate with a film thickness of 217 nm (*14*). They observed that after *c*-plane facets were formed on the surface of the α-Ga$_2$O$_3$ film, the β-Ga$_2$O$_3$ appeared with the same epitaxial relationship of ($\bar{2}$01) β-Ga$_2$O$_3$ ∥ (0001) α-Ga$_2$O$_3$/sapphire, as shown in Fig. 1B, indicating the *c*-plane facet enhanced the growth of β-Ga$_2$O$_3$ in MBE. These results suggest that a sapphire crystal plane *perpendicular* to the *c*-plane, such as *a*- or *m*-planes could potentially allow the growth of phase-pure α-Ga$_2$O$_3$ by avoiding facets, as shown in Fig. 1C. In fact, Nd-doped α-(Al$_x$Ga$_{1-x}$)$_2$O$_3$ and α-Ga$_2$O$_3$ with film thickness of 14 nm have been reported on *a*-plane sapphire in previous studies (*24*, *33*). These prior studies provided x-ray diffraction (XRD) study of the layers, but the atomic crystal structure and optical properties remain unknown.

The MBE growth of α-(Al$_x$Ga$_{1-x}$)$_2$O$_3$ material family on *m*-plane sapphire has not been investigated in detail. However, Sn-doped α-Ga$_2$O$_3$ of bandgap ~5.3 eV on *m*-plane sapphire grown by mist-CVD exhibits a room-temperature electron mobility of 65 cm$^2$/V·s at a high carrier density of $n=1.2\times10^{18}$ cm$^{-3}$ (*34*), suggesting promising electronic properties of α-(Al$_x$Ga$_{1-x}$)$_2$O$_3$ on *m*-plane sapphire. High room-temperature electron mobility of ~600 cm$^2$/V·s was concluded for optically generated electrons in α-Al$_2$O$_3$, increasing to ~4000 cm$^2$/V·s at low temperatures (*35*). A recent first-principles study predicts that Si can be an efficient shallow donor for high Al-content (Al$_x$Ga$_{1-x}$)$_2$O$_3$ (*36*), to achieve n-type conductivity of (Al$_x$Ga$_{1-x}$)$_2$O$_3$.

Motivated by the above reasons, in this study, we explore growth of this UWBG semiconductor family on *m*-plane sapphire substrate. We find that the MBE growth of α-(Al$_x$Ga$_{1-x}$)$_2$O$_3$ on *m*-plane sapphire solves the faceting problem completely, allowing α-(Al$_x$Ga$_{1-x}$)$_2$O$_3$ and its bandgap engineering over the entire composition range to be achieved. The epitaxial growths are performed by plasma-assisted MBE. The growth, and subsequent structural, chemical, and optical characterization methods are described in the Methods section.

**Results and discussion**
**SINGLE CRYSTAL ULTRA-WIDE BANDGAP EPITAXIAL LAYERS**

Symmetric XRD 2θ/ω scans of the samples are displayed in a logarithmic intensity scale in Fig. 2A. The profiles reveal that the diffraction peak is from the α-$(Al_xGa_{1-x})_2O_3$ 30$\bar{3}$0 reflection at angles slightly lower than that of the sapphire substrate. There are no peaks originating from other crystal phases or other planes. With decreasing Al composition $x$, the peaks from α-$(Al_xGa_{1-x})_2O_3$ 30$\bar{3}$0 monotonically shift to lower angles without any composition segregation, suggesting a higher Ga composition in the film. No additional peaks or fringes are observed for the MBE grown α-$Al_2O_3$ epitaxial layer, which is identical to the substrate in spite of the ~60 nm thickness, as verified by secondary ion mass spectrometry (SIMS) and discussed later. From these results, we conclude that single-phase α-$(Al_xGa_{1-x})_2O_3$ films are epitaxially stabilized successfully on $m$-plane sapphire substrates for the entire range of Al compositions without phase or composition separation/segregation.

Figure 2B shows the SIMS profiles measured on the homoepitaxial α-$Al_2O_3$ layer grown on $m$-plane sapphire. The atomic concentrations of typical impurities (H, B, C, Na, and Si) are indicated by the axis on the left, while secondary intensities of Al and O are shown on the axis on the right. The intensities of Al and O in the epitaxial layer are observed to be identical to that in the $m$-plane sapphire substrate, suggesting growth of stoichiometric $Al_2O_3$. Compared to the sapphire substrate, a large B peak is seen at the growth interface, which is a marker indicating the film thickness of the MBE grown α-$Al_2O_3$ epitaxial layer is ~60 nm. There is an increase in the Si concentration at the growth interface, whereas the H and C concentrations in the epitaxial layer remain identical to the $m$-plane sapphire bulk substrate. The profiles for H and C represent the background levels of the measurements since densities of ~$10^{18}$ cm$^{-3}$ are not expected in sapphire substrates. The Na concentration is ~$10^{16}$ cm$^{-3}$. At present we assume that the source of B is the pBN crucible used for MBE source materials, Si is from the quartz tube used as a component of the O plasma source, and Na is from the sapphire substrate. Si could accumulate during the substrate treatment using O plasma prior to growth as discussed in the Methods section.

To evaluate the strain in the epitaxial films, asymmetrical reciprocal space maps (RSMs) for 22$\bar{4}$0 reflections were taken. Figure 3A shows the relation between the (22$\bar{4}$0) and (10$\bar{1}$0) $m$-plane of the corundum structure. These results indicate that the α-$(Al_xGa_{1-x})_2O_3$ films ($0.37 \leqq x \leqq 0.74$) consist of coherent and relaxed layers, and the relaxed layers are subject to slight in-plane compressive strain [Figs. 3C-F]. On the other hand, the α-$(Al_xGa_{1-x})_2O_3$ films ($x<0.37$) were completely lattice-relaxed due to the lattice mismatches to the sapphire substrates [Fig. S1]. Because 22$\bar{4}$0 is decomposed into 30$\bar{3}$0 + $\bar{1}$2$\bar{1}$0, the directions of $Q_z$ and $Q_x$ are along 10$\bar{1}$0 (the growth direction) and $\bar{1}$2$\bar{1}$0 (one of $a$-axes), respectively. The dashed red line intersects the theoretical positions of sapphire [($Q_x$, $Q_z$)=(−0.420,0.728)] and α-Ga$_2$O$_3$ 22$\bar{4}$0 reflection [($Q_x$, $Q_z$)=(−0.401,0.695)]. The α-Al$_2$O$_3$ homoepitaxial film shows a streak along $Q_x$=−0.42 Å$^{-1}$ [Fig. 3B], which is also observed for the reflections at $Q_x$=−0.42 Å$^{-1}$ in the samples whose $x$ is larger than 0.37. The vertical streak arises from a small film thickness. Two reflection peaks of α-(Al$_x$Ga$_{1-x}$)$_2$O$_3$ are obtained at $x$ between 0.37 and 0.74: one is on the same $Q_x$ lines ($Q_x$=−0.42) as that of the sapphire substrate and the other is located between the dashed line and $Q_x$=−0.42 Å$^{-1}$ [Figs. 3C-F].

Figure 4 shows the surface atomic force microscopy (AFM) images of the *m*-plane sapphire substrate and the epitaxial α-(Al$_x$Ga$_{1-x}$)$_2$O$_3$ films. The surface morphology of the as-received substrate was extremely smooth with a root mean square (RMS) roughness as small as 0.079 nm [Fig. 4A]. The α-Al$_2$O$_3$ epitaxial film shows a small RMS roughness of 0.32 nm, but does not exhibit atomic steps, which are not observed on the starting substrate either [Fig. 4B]. Curiotto *et al.* reported a step-terrace morphology for *m*-plane sapphire substrates by annealing under Ar-O$_2$ at 1235 K for 78 hours (*37*), suggesting thermal treatment for potential improvement of the surface morphology of α-Al$_2$O$_3$ films in the future. When the Al mole fraction $x$ is smaller than 0.54, the RMS roughness remains smaller than 1.1 nm [Figs. 4E-H]. For $x$=0.59 and $x$=0.74, α-(Al$_x$Ga$_{1-x}$)$_2$O$_3$ films have rougher surface morphologies of 3.5 and 1.42 nm RMS values [Figs. 4C and D], probably due to the low substrate temperature of 650 °C.

**ELECTRON MICROSCOPY REVEALS SINGLE CRYSTALLINE UWBG LAYERS**
The XRD spectra showed the growth of single-phase α-(Al$_x$Ga$_{1-x}$)$_2$O$_3$ films on *m*-plane sapphire substrates. To investigate the atomic details of the crystal structures of the epitaxial layers, high angle annular dark field STEM (HAADF-STEM) images were taken on the cross-sectional α-(Al$_x$Ga$_{1-x}$)$_2$O$_3$ samples for $x$= 1, 0.37, 0.22, and 0. All the images were viewed along the <0001> zone axis. Figure 5 shows the cross-sectional images of the α-(Al$_x$Ga$_{1-x}$)$_2$O$_3$ film series grown on *m*-plane sapphire substrates and the corresponding enlarged images at the interfaces. As shown in Fig. 5A, since the epitaxial α-Al$_2$O$_3$ layer and the substrate have the same contrast, the exact location of the interface is invisible. Based on the corroborating SIMS measurement of Fig. 2B, an enlarged image was taken at the interface that was formed 60 nm below the surface. The uniformity of the crystal structure indicates that epitaxial α-Al$_2$O$_3$ films grow homoepitaxially on *m*-plane sapphire substrate with no visible structural defects.

On the other hand, a clear contrast between the epitaxial film and the substrate is seen for the images of the α-(Al$_{0.37}$Ga$_{0.67}$)$_2$O$_3$, α-(Al$_{0.22}$Ga$_{0.78}$)$_2$O$_3$ and α-Ga$_2$O$_3$ samples [Figs. 5B-D]. The low magnification images reveal that these films grow uniformly on *m*-plane sapphire substrates. Contrast variations inside the film are due to strain relaxation by misfit dislocations at the substrate interface, which is explained in detail in the Supplementary section [Fig. S2]. No significant phase separation or compositional segregation is observed in the epitaxial layers: they are phase pure. The film thicknesses of the α-(Al$_{0.37}$Ga$_{0.67}$)$_2$O$_3$, α-(Al$_{0.22}$Ga$_{0.78}$)$_2$O$_3$, and α-Ga$_2$O$_3$ films are measured to be 66, 57, and 60 nm, respectively, which agree well with the thicknesses (*d*) estimated from x-ray reflectivity (XRR) measurements [Table S1]. The enlarged images show that sharp heterojunction interfaces are formed between the epitaxial films and the substrates, and that the α-(Al$_x$Ga$_{1-x}$)$_2$O$_3$ ($x$≤0.37) films have an identical crystal structure and orientation as the substrate. The combination of the STEM images with XRD thus indicate single-crystalline, single phase epitaxial films are successfully grown on the *m*-plane sapphire substrate over the full 0≤$x$≤1 composition range.

**DEEP-UV SPECTROSCOPY REVEALS ULTRAWIDE BANDGAPS**

Figure 6A shows the optical transmittance spectra of the α-(Al$_x$Ga$_{1-x}$)$_2$O$_3$ films ($T$) and a bare *m*-plane sapphire substrate control sample ($T_s$) as a function of the photon energy ($hv$). Each epitaxial film shows the same high transparency in the visible and UV spectral ranges as the substrate. The bandgaps of the epitaxial α-(Al$_x$Ga$_{1-x}$)$_2$O$_3$ layers are clearly observed, in spite of their small thicknesses, indicating strong photon absorption. Note that the transmittance of α-(Al$_{0.37}$Ga$_{0.63}$)$_2$O$_3$ appears lower than the other samples because of a smaller sample size of this particular sample than the optical aperture. Each absorption coefficient ($\alpha$) was calculated using the equation $T/T_s = \exp(-\alpha d)$, and $(\alpha hv)^2$ is plotted as a function of $hv$ in Fig. 6B. When $x$=0 and 0.22, the absorption edge exhibits a double step-like onset while the α-(Al$_x$Ga$_{1-x}$)$_2$O$_3$ ($x$≥0.37) have a single step onset. The observation of double step-like onset in α-Ga$_2$O$_3$ thin films have been reported earlier, suggesting two allowed direct optical interband transitions (*13, 14*). The Ga-rich α-(Al$_x$Ga$_{1-x}$)$_2$O$_3$ films (0≤$x$≤0.37) show broader absorption onsets than the Al-rich samples, which is related to the slightly indirect bandgap character of α-Ga$_2$O$_3$ (*13*).

Direct optical bandgap energies of the α-(Al$_x$Ga$_{1-x}$)$_2$O$_3$ films were estimated from the relationship of $(\alpha hv)^2$ versus $(hv-E_g)$. As shown by the red filled circles in Fig. 6C, the experimental direct bandgaps monotonically increase as the Al composition $x$ increases. The black solid line shows the theoretical direct bandgap energies calculated by Peelaers *et al*. using hybrid density functional theory [Fig. 6D] (*38*). The experimentally measured bandgaps of the α-(Al$_x$Ga$_{1-x}$)$_2$O$_3$ epitaxial layers are for the most part in excellent agreement. A ~0.2 eV lower than the predicted values from the simple Tauc plots of $(\alpha hv)^2$ versus $(hv-E_g)$ may be due to an underestimation of the bandgap energy due to excitonic absorption with low energy tails, related to the slightly indirect character of α-Ga$_2$O$_3$, and to as-yet unknown defects (*13*). The bowing parameter ($b$) is obtained from $E_g(x) = (1-x)E_g[\text{Ga}_2\text{O}_3] + xE_g[\text{Al}_2\text{O}_3] - bx(1-x)$, where $E_g[\text{Ga}_2\text{O}_3]$ and $E_g[\text{Al}_2\text{O}_3]$ are the bandgap energies of α-Ga$_2$O$_3$ and α-Al$_2$O$_3$, by fitting the plot as depicted by the red dashed line. The value of $b$ is estimated to be 1.1 eV. This experimental value agrees well with the theoretically calculated value of 1.37 eV, and proves the ultra-wide bandgap nature of the epitaxial layers available for bandgap engineering from 5.4 - 8.6 eV.

**Conclusion**

In conclusion, we have demonstrated the successful epitaxial stabilization of ultra-wide bandgap single phase α-(Al$_x$Ga$_{1-x}$)$_2$O$_3$ films over the entire composition range on *m*-plane sapphire substrates by MBE by avoiding phase transformation by the choice of crystal orientation for epitaxy. Cross-sectional HAADF-STEM images reveal single-phase, single crystalline α-(Al$_x$Ga$_{1-x}$)$_2$O$_3$ epitaxial films are formed, by avoiding the formation of *c*-plane facets. By varying the alloy composition, bandgap energies from ~5.4 eV up to 8.6 eV with a bowing parameter of 1.1 eV are achieved, making α-(Al$_x$Ga$_{1-x}$)$_2$O$_3$ the largest bandgap epitaxial material family to date. If these layers can be controllably doped, it would pave the way for α-(Al$_x$Ga$_{1-x}$)$_2$O$_3$-based high-power heterostructure electronic and photonic devices at bandgaps far beyond all materials available today.

**Materials and Methods**
**Materials**
All films reported in this work were grown on double-side polished *m*-plane (10$\bar{1}$0) sapphire substrates in a Veeco Gen930 MBE system equipped with standard effusion cells for elemental Ga and Al, and a radio frequency (RF) plasma source for active oxygen species. The pressure in the growth chamber was ~10$^{-6}$ Torr during the growth runs. For the growth of α-(Al$_x$Ga$_{1-x}$)$_2$O$_3$ films ($x$≥0.37), the input RF power and oxygen flow rate were fixed at 250 W and 0.50 sccm, respectively. The α-Al$_2$O$_3$ and α-(Al$_x$Ga$_{1-x}$)$_2$O$_3$ films (1>$x$≥0.37) were grown for 2 hours at thermocouple substrate temperatures ($T_{sub}$) of 750 and 650 °C, respectively. The α-(Al$_x$Ga$_{1-x}$)$_2$O$_3$ ($x$=0.22 and 0) samples were grown at $T_{sub}$=650 °C using the input RF power of 260 W and the oxygen flow rate of 0.23 sccm. The growth time of α-(Al$_{0.22}$Ga$_{0.78}$)$_2$O$_3$ and α-Ga$_2$O$_3$ were 3 hours, and 3 hours 13 minutes, respectively. The Al concentration ($x$) in the films was varied from 0 to 1 by regulating the beam equivalent pressures (BEPs) of Ga [(0–1.1) ×10$^{-8}$ Torr] and Al [(0-1.6) ×10$^{-8}$ Torr]. Before each growth, the *m*-plane sapphire substrates were subjected to an oxygen plasma treatment in the growth chamber for 10 min at $T_{sub}$=800 °C.

**Characterization of materials**
Structural properties and surface morphology of the samples were characterized by XRD measurements, AFM, and STEM. The film thicknesses of the samples ($x$<1) were determined by XRR measurements, while SIMS was used to obtain the chemical impurity species as well as the thickness of the α-Al$_2$O$_3$ epitaxial layer, since there were no thickness fringes observed from the XRR profile for the α-Al$_2$O$_3$ film due to homoepitaxial growth. The film thicknesses were in the range of 56–84.3 nm. Table S1 summarizes the Al and Ga BEPs and film thicknesses ($d$) of the samples. The Al composition was estimated by XRD and x-ray photoelectron spectroscopy (XPS). Optical bandgaps of the films were estimated by analyzing the optical transmission spectra; a *m*-plane sapphire substrate was used as a reference sample. The optical transmittance spectra were measured at room temperature (RT) using a Varian Cary 50 UV-Vis spectrometer with a wavelength range of 190–800 nm for $x$≤0.22 and using a Bunkoukeiki KV-201 vacuum UV spectrophotometer with a wavelength range of 140–300 nm for $x$≥0.37. The Bunkoukeiki KV-201 unit had a 20-cm focal-length Czerny-Turner monochromator equipped with a 1200 groove/mm grating and a 30 W deuterium lamp that were purged with nitrogen gas during operation. A separate α-Al$_2$O$_3$ sample used for transmittance measurement was grown at $T_{sub}$=750 °C for 2 hours using at the input RF power of 250 W, the oxygen flow rate of 0.50 sccm, and Al BEP of 1.5×10$^{-8}$

Torr. By inserting a thin α-Ga$_2$O$_3$ layer to intentionally introduce thickness fringes, the film thickness of 78.2 nm was measured by XRR. We used the sample directly grown on sapphire in the same growth condition for the transmittance measurement.

**Transmission electron microscopy measurement**

Cross-sectional TEM specimens were prepared using a FEI Strata 400 Focused Ion Beam (FIB) with a final milling step of 5 keV to reduce damage. Carbon and platinum protective layers were deposited prior to milling in order to minimize ion-beam damage. The samples were then examined by STEM, using an aberration corrected Titan Themis operating at 300 keV.

**Acknowledgments**
The authors acknowledge useful discussions with Michael Thompson, Farhan Rana, and Nicholas Tanen of Cornell University, and Kelson Chabak, Andrew Neal, Andrew Green, Shin Mou, and Thaddeus Asel of the Air Force Research Laboratories.

**Funding:** This work was in part supported by JSPS Overseas Challenge Program for Young Researchers 1080033, and by the Air-Force/Cornell Center for Epitaxial Solutions (ACCESS) center of excellence monitored by Dr. Ali Sayir.

**Author contributions:** R.J., Y.C., and D.J. conceived the study and designed the experimental study. R.J., C.C., T.O., S.H., M.C.C., K.L., performed the experiments. R.J., C.C., T.O., Y.C., and conducted the subsequent data analysis. R.J., C.C., and D.J. wrote the manuscript. V.V., D.G.S., D.A.M., and H.G.X. contributed ideas and feedback on the analysis and edited the manuscript. All authors edited the manuscript and approved its final version.

**Competing interests:** The authors declare that they have no competing interests.

**Data and materials availability:** All data needed to evaluate the conclusions in the paper are present in the paper and/or the Supplementary Materials. Additional data related to this paper may be requested from the authors

**Figures and Tables**

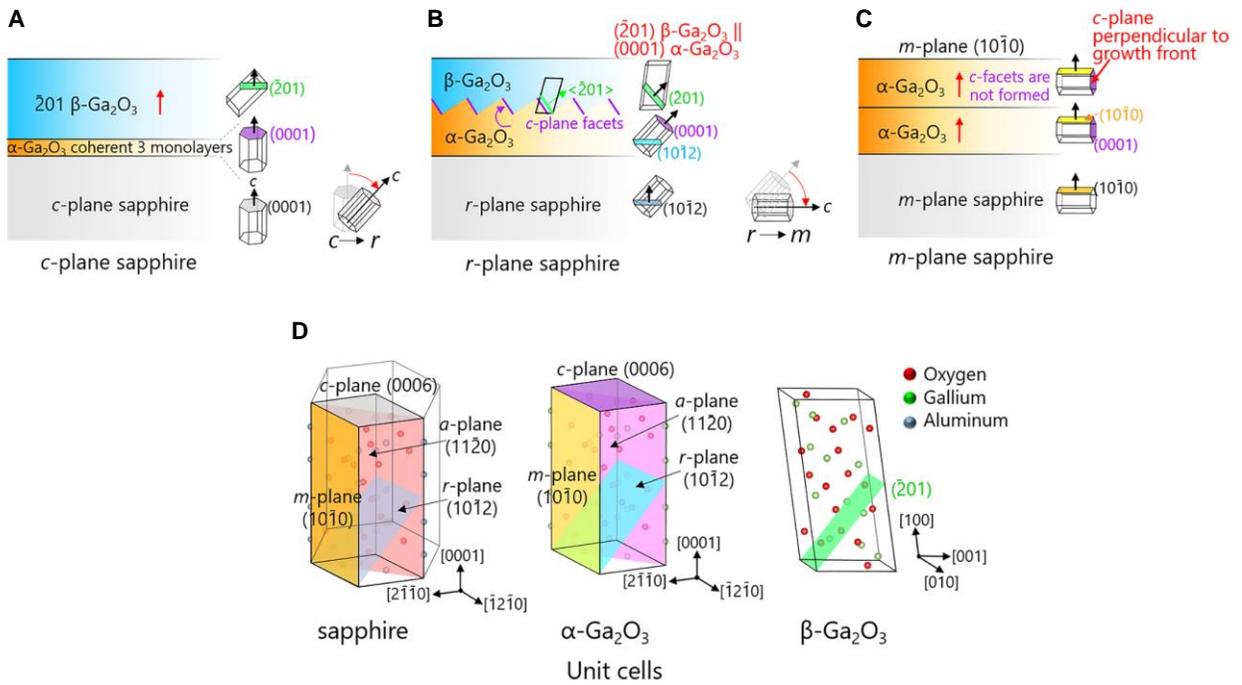

**Fig. 1. Schematics of growth behaviors of Ga$_2$O$_3$ grown on sapphire.** Schematics of growth behaviors of Ga$_2$O$_3$ grown on (**A**) *c*-, (**B**) *r*- and (**C**) *m*-plane sapphire substrates by MBE, and (**D**) unit cells of sapphire, α-Ga$_2$O$_3$ and β-Ga$_2$O$_3$. β-Ga$_2$O$_3$ is grown on *c*-plane sapphire substrates with the coherent α-Ga$_2$O$_3$ interlayer. [32] The epitaxial relationship is ($\bar{2}$01) β-Ga$_2$O$_3$ ∥ (0001) α-Ga$_2$O$_3$/sapphire. On *r*-plane sapphire, after *c*-plane facets are formed on the surface of the α-Ga$_2$O$_3$ layer, β-Ga$_2$O$_3$ appears with the same epitaxial relationship of ($\bar{2}$01) β-Ga$_2$O$_3$ ∥ (0001) α-Ga$_2$O$_3$. [14] On the other hand, *m*-plane, which is perpendicular to the *c*-plane, is expected to have less chance to form the *c*-plane facets and allow the growth of phase-pure α-Ga$_2$O$_3$.

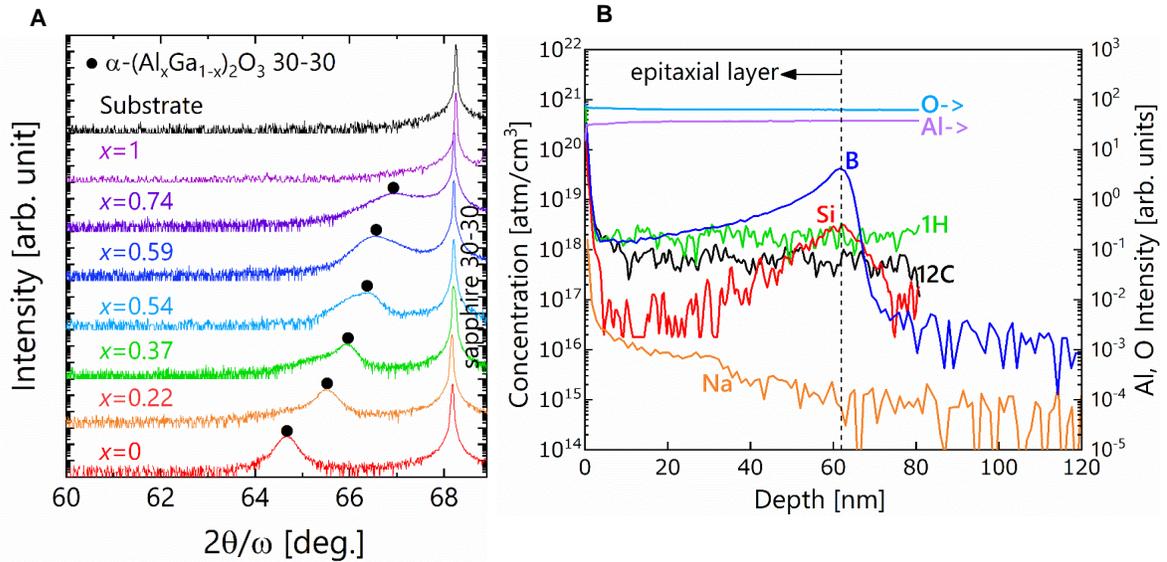

**Fig. 2. X-ray diffraction scans of α-(Al$_x$Ga$_{1-x}$)$_2$O$_3$ films and SIMS profiles of the α-Al$_2$O$_3$ epitaxial film.** (**A**) Symmetric XRD 2θ/ω scans of α-(Al$_x$Ga$_{1-x}$)$_2$O$_3$ films grown on *m*-plane sapphire substrates by plasma-assited MBE. Diffraction peaks from the α-(Al$_x$Ga$_{1-x}$)$_2$O$_3$ are denoted by the filled circles. Note that no diffraction peaks from other phases are observed. (**B**) Atomic densities of typical impurities in the α-Al$_2$O$_3$ epitaxial film grown on *m*-plane sapphire measured by SIMS. The film/substrate interface is marked by the clear peak in the B profile, indicating the epitaxial film thickness is ~60 nm. The concentrations of H and C are at background levels. Secondary ion intensities of Al and O are also shown on the side axis on the right, indicating stoichiometric epitaxial growth.

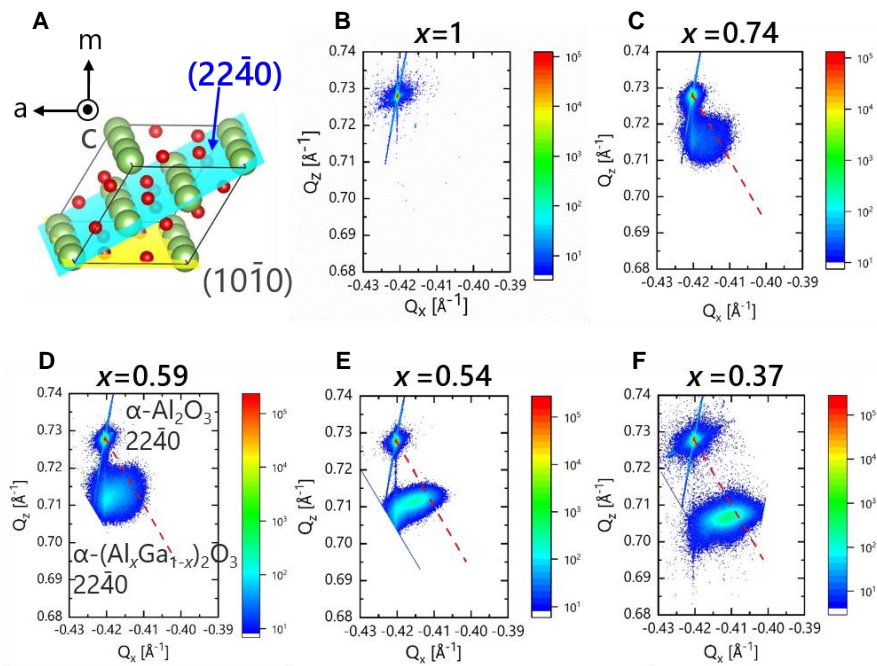

**Fig. 3. Reciprocal space maps ($Q_x$, $Q_z$) around the $22\bar{4}0$ reflections of the α-(Al$_x$Ga$_{1-x}$)$_2$O$_3$ epitaxial films.** (**A**) The relation between the ($22\bar{4}0$) and ($10\bar{1}0$) *m*-plane of the corundum structure. The directions of $Q_z$ and $Q_x$ are along $10\bar{1}0$ and $\bar{1}2\bar{1}0$, respectively. Reciprocal space maps for *x*= (**B**) 0, (**C**) 0.74, (**D**) 0.59, (**E**) 0.54, and (**F**) 0.37.

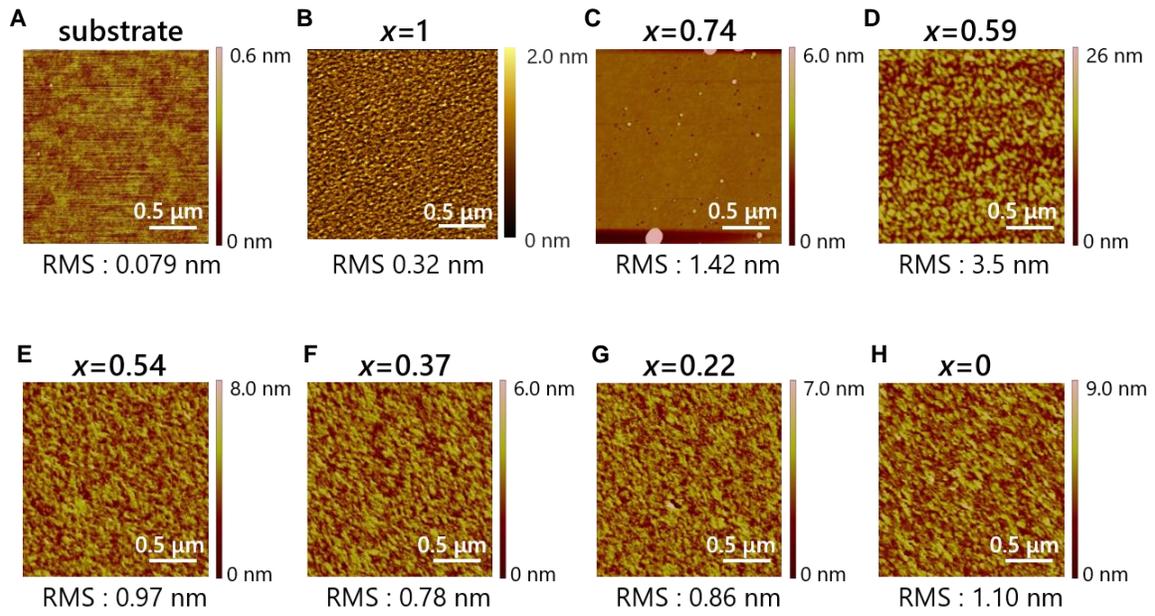

**Fig. 4. Surface morphology of α-(Al$_x$Ga$_{1-x}$)$_2$O$_3$ films on *m*-plane sapphire substrates.** (**A**) sapphire substrate. (**B**)-(**H**) are α-(Al$_x$Ga$_{1-x}$)$_2$O$_3$ films with *x*=1, 0.74, 0.59, 0.54, 0.37, 0.22, and 0, respectively. The RMS roughness of the α-(Al$_x$Ga$_{1-x}$)$_2$O$_3$ films for *x*≤0.54 is less than 1.1 nm.

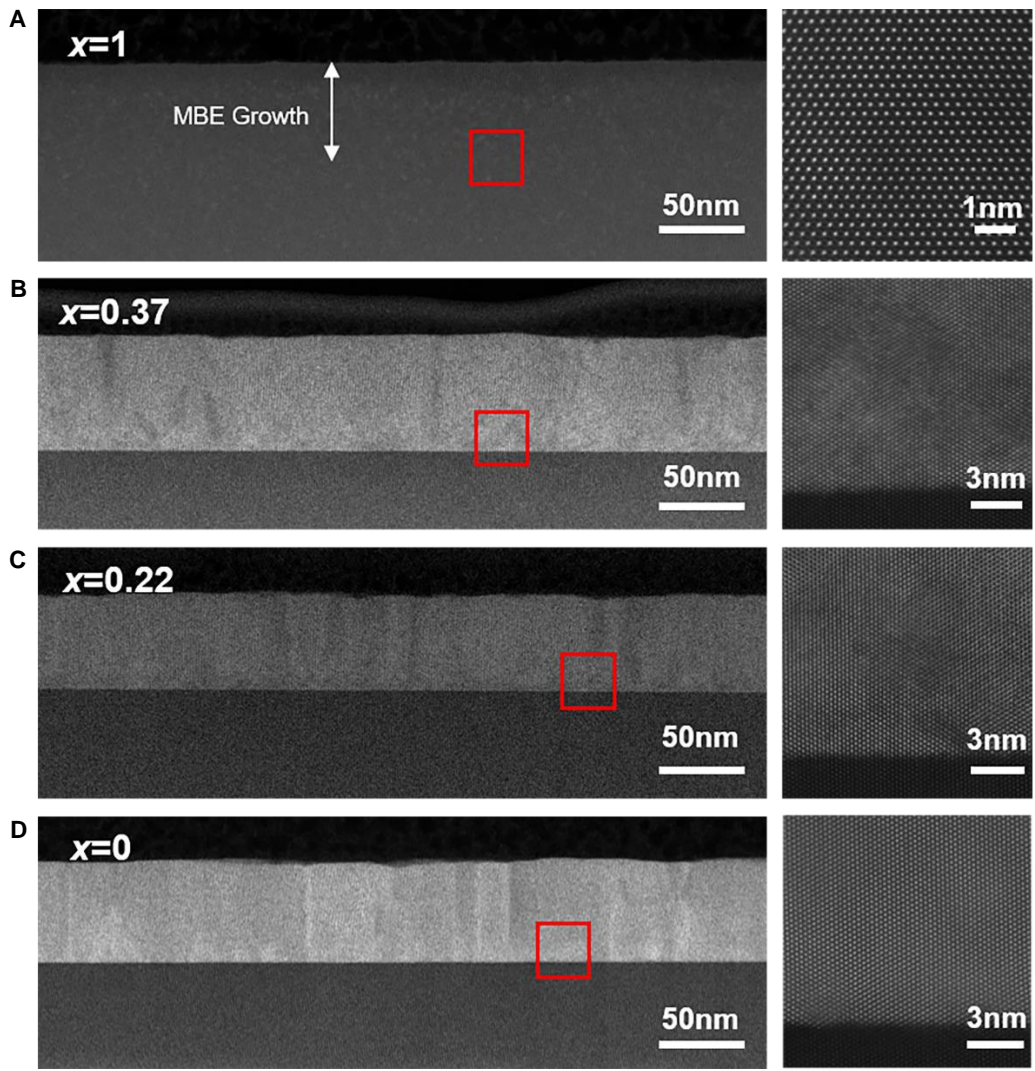

**Fig. 5. HAADF-STEM images showing an overview of α-(Al$_x$Ga$_{1-x}$)$_2$O$_3$ film grown on *m*-plane sapphire and the enlarged interfaces from the boxes in the overview images.** (**A**)-(**D**) corresponds to *x*=1, *x*=0.37, *x*=0.22, *x*=0 respectively. (**A**) The film and the substrate show identical contrast, as expected for a homoepitaxial film. The enlarged area shows no differences in contrast and lattice structure. (**B**), (**C**), (**D**) show film thickness of 66nm, 55nm, and 60nm respectively. They all show sharp interfaces while the films show contrast variations from strain relaxation resulting from misfit dislocations at the interface. Further details of the defects are explained in the Supplementary section. Enlarged atomic resolution image in (**D**) shows that α-Ga$_2$O$_3$ film is relaxed at the interface.

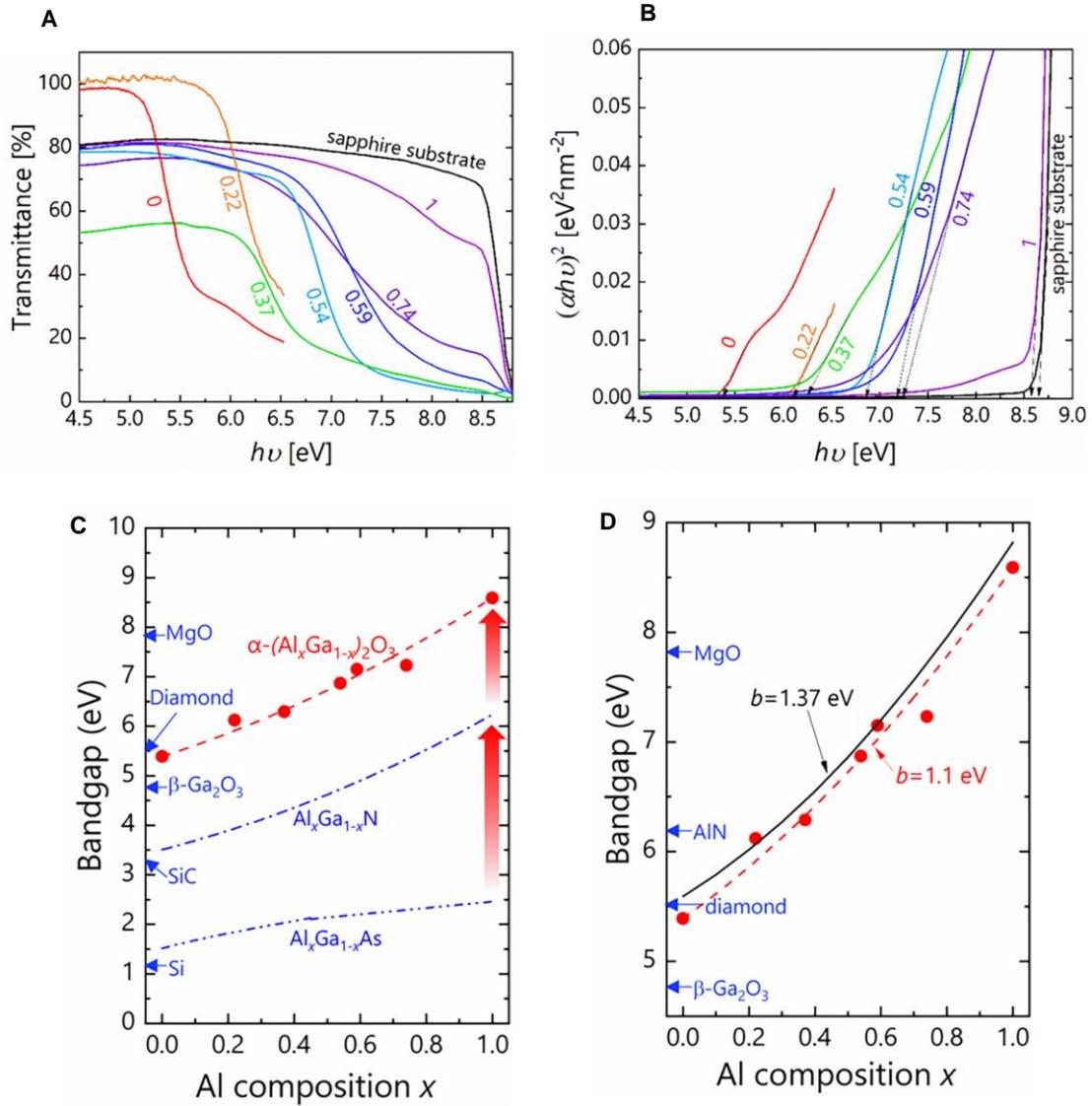

**Fig. 6. Optical transmittance spectra and bandgap energies of the α-(Al$_x$Ga$_{1-x}$)$_2$O$_3$ films.** (**A**) External optical transmittance of the α-(Al$_x$Ga$_{1-x}$)$_2$O$_3$ films and substrate, (**B**) $(\alpha h\nu)^2$-$h\nu$ plot as a function of the Al composition. The α-(Al$_{0.37}$Ga$_{0.63}$)$_2$O$_3$ film exhibited a lower transmittance than the other samples since the sample was smaller than the aperture for the transmittance measurement. Direct bandgaps (solid symbols) for the α-(Al$_x$Ga$_{1-x}$)$_2$O$_3$ films as a function of the Al composition shown in (**C**) the wide range from 0 to 10 eV and (**D**) the ultra-wide bandgap region ($E_g$>4 eV). The red solid line is a quadratic fit to the direct bandgaps. The black dashed line is quadratic fits to the computed direct bandgap for the corundum structure as reported in ref 38). The bandgaps for AlGaN and AlGaAs are also shown for comparison. [39]

**Supplementary Materials**

**Reciprocal Space Maps of α-Ga$_2$O$_3$ and α-(Al$_{0.18}$Ga$_{0.82}$)$_2$O$_3$ on *m*-plane Sapphire Substrates**

Asymmetric RMSs were observed for α-Ga$_2$O$_3$ and α-(Al$_{0.18}$Ga$_{0.82}$)$_2$O$_3$ epitaxial films grown on *m*-plane sapphire substrates. The films were grown for 2 hours at $T_{sub}$ of 650 °C using a RF power of 250 W and oxygen flow rate of 0.50 sccm. The Ga BEP was $1.1\times10^{-8}$ Torr for the α-Ga$_2$O$_3$ while the BEPs of Al and Ga were set at $2.5\times10^{-9}$ and $9\times10^{-9}$ Torr, respectively, for the α-(Al$_{0.18}$Ga$_{0.82}$)$_2$O$_3$. The film thicknesses of the α-Ga$_2$O$_3$ and α-(Al$_{0.18}$Ga$_{0.82}$)$_2$O$_3$ determined by XRR measurements were 52 and 57 nm. As shown in Fig. S1, both α-Ga$_2$O$_3$ and α-(Al$_{0.18}$Ga$_{0.82}$)$_2$O$_3$ films, though single-crystalline, are lattice-relaxed due to the large lattice mismatch to the sapphire substrates. This is similar to the growth of GaN on SiC, Sapphire, or Silicon.

**Observation of Defects in α-Ga$_2$O$_3$ by TEM Imaging**

To further confirm that the samples grown are single phase, we performed selected area electron diffraction (SAED) imaging using FEI T-12 operating at 120 kV. Figure S2 shows an example of the epitaxial α-Ga$_2$O$_3$ sample. The electron diffraction pattern for the α-Ga$_2$O$_3$/*m*-plane sapphire shows two 'concentric' hexagonal patterns [Fig. S2(a)]. The lattice constant of sapphire is smaller than that of α-Ga$_2$O$_3$, resulting in the hexagonal diffraction pattern of the sapphire being slightly further from the center. In Fig. S2(b), the dark-field TEM image is color-coded in red and blue, where each color shows the signals coming from the inner α-Ga$_2$O$_3$ (circled in red) and outer sapphire diffraction spot (circled in blue) from Fig. S2(a). The α-Ga$_2$O$_3$ epitaxial layer is not completely red, because strain has distorted the lattice and therefore some regions in the film are not perfectly on the same crystal zone axis.

HAADF-STEM images at the α-Al$_2$O$_3$/α-Ga$_2$O$_3$ heterojunction show that the α-Ga$_2$O$_3$ layer is relaxed at the interface through the introduction of misfit dislocations [Fig. S2(c)]. In addition, some isolated areas near the substrate interface exhibit a crystal structure different from α-Ga$_2$O$_3$ [Fig. S2(d)]. In contrast to the α-Ga$_2$O$_3$ viewed down along <0001> zone axis, the defective region displays a distinctive pattern of edge-sharing rings. The schematic explains how such defective regions can form. When α-Ga$_2$O$_3$ is viewed along the <0001> zone axis, the Ga atoms arrange in a traditional hexagonal-closed-packed (hcp) structural form, as shown in the green atomic lattice structure. Here the following displacements occur: the yellow crystal structure, identical to the green, glides half a unit cell along the <11$\bar{2}$0> axis (horizontal direction). As HAADF-STEM images are obtained by averaging through several unit cells in projection, such a glide can result in the overlap seen in the lattice structure in the image resulting in the observed chained hexagonal lattice regions. This displacement is found to occur in different glide directions and distances as well, resulting in distinctive lattice structures in some isolated regions in the film. This stacking fault seems to be caused by the strain at the interface. This defect was also observed in the low Al-content α-(Al$_{0.22}$Ga$_{0.78}$)$_2$O$_3$ films as well. For α-(Al$_{0.37}$Ga$_{0.67}$)$_2$O$_3$, we observed loss of focus inside the film, which we believe is potentially caused by small columnar defect regions that are obscured by the thickness of the film. Although the dislocations and the stacking faults exist at the heterointerface in the high Ga-content α-(Al$_x$Ga$_{1-x}$)$_2$O$_3$ layers, STEM images indicate single-crystalline, single phase epitaxial films are successfully grown on the *m*-plane sapphire substrate over the full 0≤*x*≤1 composition range. No defects are observable in STEM in the α-Al$_2$O$_3$ epitaxial layer.

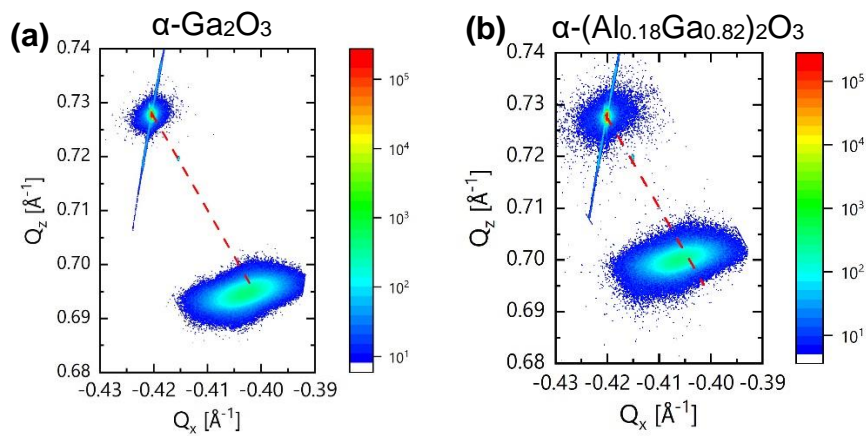

**Fig. S1.** Reciprocal space maps ($Q_x$, $Q_z$) around the 22$\bar{4}$0 reflections of the (**b**) α-Ga$_2$O$_3$ and (**a**) α-(Al$_{0.18}$Ga$_{0.82}$)$_2$O$_3$ epitaxial films.

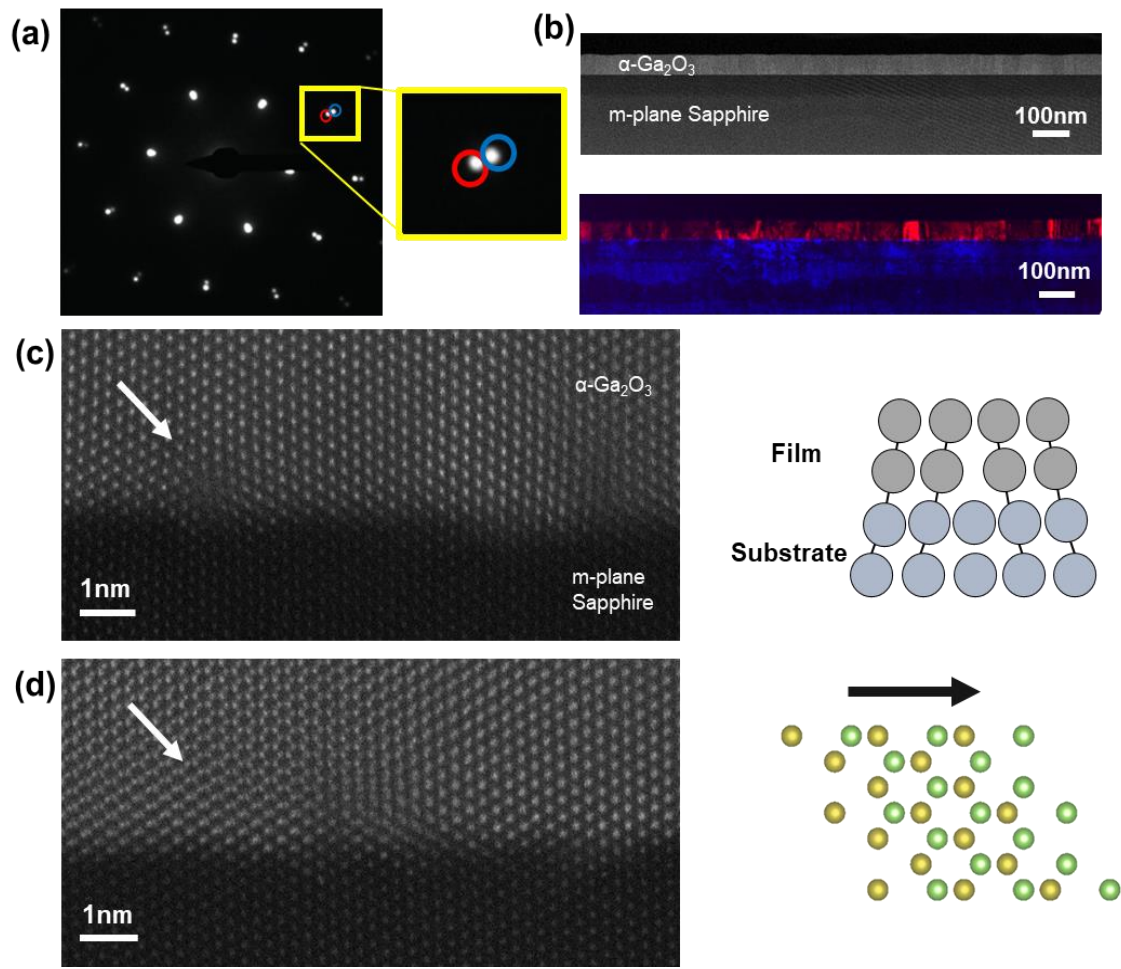

**Fig. S2. Cross-sectional STEM images of the α-(Al$_x$Ga$_{1-x}$)$_2$O$_3$ film with $x$=0.** (a) Diffraction yields two similar hexagonal patterns each originating from α-Ga$_2$O$_3$ and sapphire. As the lattice constant of sapphire is slightly smaller than that of α-Ga$_2$O$_3$, the outer diffraction spot of the two adjacent spots shown (circled in blue) corresponds to sapphire as shown in (b). The inner spot (circled in red) corresponds to the α-Ga$_2$O$_3$ film. False-color dark-field image overlay corresponding to each diffraction spots circled in blue and red are shown in (b), together with a HAADF-STEM image. The dark-field image of the α-Ga$_2$O$_3$ film is not completely color-coded in red as the strain has distorted the lattice. (c) Enlarged epitaxial interface reveals the atomic structure that is relaxing the strain. The schematic shows the corresponding edge dislocation. (d) Defective areas showing different hexagonal crystal structure from that of α-Ga$_2$O$_3$ that is caused by strain at the interface. A schematic of α-Ga$_2$O$_3$ crystal structure is also shown to explain how such hexagonal crystal lattices can form. When the yellow α-Ga$_2$O$_3$ crystal structure glides half a unit cell, the resulting crystal structure appears as a chained hexagonal structure when seen in projection in TEM.

**Table S1.** The Al and Ga BEPs and film thicknesses of the α-(Al$_x$Ga$_{1-x}$)$_2$O$_3$ films grown on *m*-plane sapphire substrates.

| $x$ | 0 | 0.22 | 0.37 | 0.54 | 0.59 | 0.74 | 0 |
|---|---|---|---|---|---|---|---|
| Al BEP (10$^{-9}$ Torr) | 0 | 1.5 | 4.9 | 9 | 9 | 9 | 16 |
| Ga BEP (10$^{-9}$ Torr) | 9 | 7.5 | 9 | 9 | 4.7 | 3 | 0 |
| Film thickness $d$ (nm) | 57 | 56 | 66 | 84.3 | 63.8 | 58.1 | ~60 |